\documentclass[
reprint,
superscriptaddress,
 amsmath,amssymb,
 aps,prl
]{revtex4-2}

\usepackage{graphicx}
\usepackage{dcolumn}
\usepackage{bm}
\usepackage{xcolor}
\usepackage{changes}
\usepackage{url}
\usepackage{hyperref}
\usepackage[mathlines]{lineno}

\newcommand{\NUCLEUS}{\textsc{Nucleus} }
\newcommand{\FIFRELIN}{\textsc{FIFRELIN}}

\newcommand{\CRESST}{\textsc{Cresst}}

\newcommand{\GEANT}{\textsc{GEANT4} }
\newcommand{\CEvNS}{\textsc{CE$\nu$NS}}
\newcommand{\CRAB}{\textsc{CRAB}}

\newcommand{\nuc}[2] {$^{\text{#1}}$#2}

\begin{document}

\preprint{APS/123-QED}
\title{Observation of a nuclear recoil peak at the 100\,eV scale induced by neutron capture
}
 
\newcommand{\CEACadarache}{\affiliation{CEA, DES, IRESNE, DER, Cadarache F-13108 Saint-Paul-Lez-Durance, France}}
\newcommand{\CEASaclay}{\affiliation{IRFU, CEA, Universit\'e Paris-Saclay, 91191 Gif-sur-Yvette, France}}
\newcommand{\CEASRMP}{\affiliation{CEA, DES, SRMP, Universit\'e Paris-Saclay, 91191 Gif-sur-Yvette, France}}
\newcommand{\LNGS}{\affiliation{Istituto Nazionale di Fisica Nucleare -- Laboratori Nazionali del Gran Sasso, 67100 Assergi, Italy}}
\newcommand{\MPP}{\affiliation{Max-Planck-Institut für Physik, D-80805 München, Germany}}
\newcommand{\TUM}{\affiliation{Physik-Department, Technische Universität München, D-85748 Garching, Germany}}
\newcommand{\TUW}{\affiliation{Atominstitut, Technische Universität Wien, A-1020 Wien, Austria}}
\newcommand{\IJCL}{\affiliation{Universit\'e Paris-Saclay, CNRS/IN2P3, IJCLab, 91405 Orsay, France}}

\newcommand{\APC}{\affiliation{APC, Universit\'{e} de Paris, CNRS, Astroparticule et Cosmologie, Paris F-75013, France}}

\newcommand{\HEPHY}{\affiliation{Institut f\"ur Hochenergiephysik der \"Osterreichischen Akademie der Wissenschaften, A-1050 Wien, Austria}}

\newcommand{\Sapienza}{\affiliation{Dipartimento di Fisica, Sapienza Universit\`{a} di Roma, Roma I-00185, Italy}}

\newcommand{\TorVergata}{\affiliation{Dipartimento di Fisica, Universit\`{a} di Roma ``Tor Vergata", Roma I-00133, Italy}}

\newcommand{\Ferrara}{\affiliation{Dipartimento di Fisica, Universit\`{a} di Ferrara, I-44122 Ferrara, Italy}}

\newcommand{\INFNRoma}{\affiliation{Istituto Nazionale di Fisica Nucleare -- Sezione di Roma, Roma I-00185, Italy}}

\newcommand{\INFNTorVergata}{\affiliation{Istituto Nazionale di Fisica Nucleare -- Sezione di Roma ``Tor Vergata", Roma I-00133, Italy}}

\newcommand{\INFNFerrara}{\affiliation{Istituto Nazionale di Fisica Nucleare -- Sezione di Ferrara, I-44122 Ferrara, Italy}}

\newcommand{\CNR}{\affiliation{Consiglio Nazionale delle Ricerche, Istituto di Nanotecnologia, Roma I-00185, Italy}}

\newcommand{\CIUC}{\affiliation{LIBPhys-UC, Departamento de Fisica, Universidade de Coimbra, P3004 516 Coimbra, Portugal}}

\author{H.~Abele}\TUW
\author{G.~Angloher}\MPP
\author{A.~Bento}\MPP\CIUC
\author{L.~Canonica}\MPP
\author{F.~Cappella}\INFNRoma
\author{L.~Cardani}\INFNRoma
\author{N.~Casali}\INFNRoma
\author{R.~Cerulli}\INFNTorVergata\TorVergata
\author{A.~Chalil}\CEASaclay
\author{A.~Chebboubi}\CEACadarache
\author{I.~Colantoni}\INFNRoma\CNR
\author{J.-P.~Crocombette}\CEASRMP
\author{A.~Cruciani}\INFNRoma
\author{G.~Del~Castello}\INFNRoma\Sapienza
\author{M.~del~Gallo~Roccagiovine}\INFNRoma\Sapienza
\author{D.~Desforge}\CEASaclay
\author{A.~Doblhammer}\TUW
\author{E.~Dumonteil}\CEASaclay
\author{S.~Dorer}\TUW
\author{A.~Erhart}\TUM
\author{A.~Fuss}\TUW\HEPHY
\author{M.~Friedl}\HEPHY
\author{A.~Garai}\MPP
\author{V.~M.~Ghete}\HEPHY
\author{A.~Giuliani}\IJCL
\author{C.~Goupy}\CEASaclay
\author{F.~Gunsing}\CEASaclay
\author{D.~Hauff}\MPP
\author{F.~Jeanneau}\CEASaclay
\author{E.~Jericha}\TUW
\author{M.~Kaznacheeva}\TUM
\author{A.~Kinast}\TUM
\author{H.~Kluck}\HEPHY
\author{A.~Langenk\"{a}mper}\MPP 
\author{T.~Lasserre}\CEASaclay\TUM
\author{A.~Letourneau}\CEASaclay
\author{D.~Lhuillier}\CEASaclay
\author{O.~Litaize}\CEACadarache
\author{M.~Mancuso}\MPP
\author{P.~de Marcillac}\IJCL
\author{S.~Marnieros}\IJCL
\author{T.~Materna}\CEASaclay
\author{B.~Mauri}\CEASaclay
\author{A.~Mazzolari}\INFNFerrara
\author{E.~Mazzucato}\CEASaclay
\author{H.~Neyrial}\CEASaclay
\author{C.~Nones}\CEASaclay
\author{L.~Oberauer}\TUM
\author{T.~Ortmann}\TUM
\author{A.~Ouzriat}\CEASaclay 
\author{L.~Pattavina}\TUM\LNGS
\author{L.~Peters}\TUM
\author{F.~Petricca}\MPP
\author{D.~V.~Poda}\IJCL
\author{W.~Potzel}\TUM
\author{F.~Pr\"{o}bst}\MPP
\author{F.~Reindl}\TUW\HEPHY
\author{R.~Rogly}\CEASaclay
\author{M.~Romagnoni}\INFNFerrara
\author{J.~Rothe}\TUM
\author{N.~Schermer}\TUM
\author{J.~Schieck}\TUW\HEPHY
\author{S.~Sch\"{o}nert}\TUM
\author{C.~Schwertner}\TUW\HEPHY
\author{L.~Scola}\CEASaclay
\author{O.~Serot}\CEACadarache
\author{G.~Soum-Sidikov}\CEASaclay
\author{L.~Stodolsky}\MPP
\author{R.~Strauss}\TUM
\author{M.~Tamisari}\INFNFerrara\Ferrara
\author{L.~Thulliez}\CEASaclay
\author{C.~Tomei}\INFNRoma
\author{M.~Vignati}\INFNRoma\Sapienza
\author{M.~Vivier}\CEASaclay
\author{V.~Wagner}\TUM
\author{A.~Wex}\TUM

\collaboration{\CRAB{} Collaboration}
\email{correspondence: crab.neutrons@gmail.com}
\collaboration{\NUCLEUS{}Collaboration}
\email{correspondence: info@nucleus-experiment.org}
\noaffiliation

\date{\today}

\begin{abstract}
Coherent elastic neutrino-nucleus scattering and low-mass Dark Matter detectors rely crucially on the understanding of their response to nuclear recoils. We report the first observation of a  nuclear recoil peak at around 112\,eV induced by neutron capture. The measurement was performed with a CaWO$_4$ cryogenic detector from the NUCLEUS experiment exposed to a \nuc{252}{Cf} source placed in a compact moderator. We identify the expected peak structure from the single-$\gamma$ de-excitation of \nuc{183}{W} with 3\,$\sigma$ and its origin by neutron capture with 6\,$\sigma$ significance. 
This result demonstrates a new method for precise, in-situ, and non-intrusive calibration of low-threshold experiments. 
\end{abstract}

\maketitle


Cryogenic detectors have demonstrated extremely low energy thresholds down to a few tens of eV, with detector masses of the order of 1 to 100\,g~\cite{Strauss:2017cam, CRESST:2019jnq,PhysRevD.99.082003,EDELWEISS:2022ktt,SuperCDMS:2020aus,SuperCDMS:2018mne}. These achievements enable a wide experimental program: the extension of Dark Matter (DM) searches to particles with masses below 1\,GeV/c$^2$~\cite{Angloher:2017sxg,SuperCDMS:2018mne}, and the exploration of coherent elastic neutrino-nucleus scattering (\CEvNS{})~\cite{Freedman:1973yd,Drukier:1983gj,Abdullah:2022zue}. Access to this new neutrino interaction channel will enable complementary tests of the Standard Model~\cite{AristizabalSierra:2021uob} at low energies and searches for new physics~\cite{Lindner:2016wff,Dent:2016wcr}. 
The \NUCLEUS{} experiment~\cite{Strauss:2017cuu,Rothe:2019aii,Angloher:2019flc} is one of several experiments under development worldwide to study \CEvNS{} at nuclear reactors~\cite{Ricochet:2021rjo,Agnolet:2016zir,bonet2021constraints,CONNIE:2019xid,Akimov:2017hee}. A major challenge is the precise calibration of the  nuclear recoil signature at the 10--100\,eV scale due to the lack of suitable low-energy calibration sources. The DM and \CEvNS{} communities prioritize the development of calibration techniques in the so-called ``UV-gap" at 12--50\,eV \cite{PhysRevD.102.063026,Baxter:2022dkm}, above the energy of VUV photons and below the reach of X-ray fluorescence sources. Of particular interest is the direct calibration of low-energetic nuclear recoils, which is the experimental signature of DM and \CEvNS{}. At energy depositions of $\leq$\,100\,eV, the energy stored in lattice defects becomes comparable to the total deposited energy and  crystal-defect creation increasingly impacts the visible (phonon-mediated) energy~\cite{PhysRevD.106.063012}. This effect is not accessible with state-of-the-art calibration techniques relying on electron recoils and requires the measurement of known nuclear recoils.

The capture of thermal neutrons in crystals has been recently proposed to provide an in-situ calibration by inducing peaks of nuclear recoils as low as a few tens of eV, uniformly distributed in the volume of the detector.  
This technique was first suggested by the CRAB (Calibrated nuclear Recoils for Accurate Bolometry) collaboration in~\cite{Thulliez:2020esw}. 
We describe the first observation of a neutron-capture-induced nuclear recoil peak using a low-threshold CaWO$_4$ \NUCLEUS{} cryogenic detector~\cite{Strauss:2017cam}. The method can be applied to different detector materials such as Ge~\cite{Thulliez:2020esw}, and a first experimental attempt to observe the process was performed using a Si cryogenic detector~\cite{Villano:2021eof}.

For CaWO$_4$ the most prominent expected calibration feature is a peak at 112.5\,eV~\cite{Thulliez:2020esw}.
It is associated with the nuclear recoil induced by the single-$\gamma$ de-excitation of \nuc{183}{W} after capturing a neutron on \nuc{182}{W} which has a natural abundance of 26.50\% in tungsten~\cite{isoAbundance}. The corresponding $\gamma$ has an energy of 6.191\,MeV and given an attenuation of 3.7$\cdot10^{-2}$\,cm$^2$/g in CaWO$_4$~\cite{NISTxcom} such $\gamma$-rays have a high probability to escape a gram-scale crystal. The resulting energy deposition in the detector is hence a pure  nuclear recoil signal. 
The single-$\gamma$ emission competes with the de-excitation by multi-$\gamma$ cascades, causing a broad distribution of  nuclear recoils on the left side of the single-$\gamma$ recoil peak~\cite{Thulliez:2020esw}. 
The neutron capture on \nuc{182}{W} is the dominant process with a cross-section of 20.32\,barn~\cite{Brown:2018jhj}. Capture reactions on other W isotopes yield less prominent peaks at energies of 85 and 160\,eV~\cite{Thulliez:2020esw}.

\begin{figure}[t]
    \centering
    \includegraphics[width=0.95\linewidth]{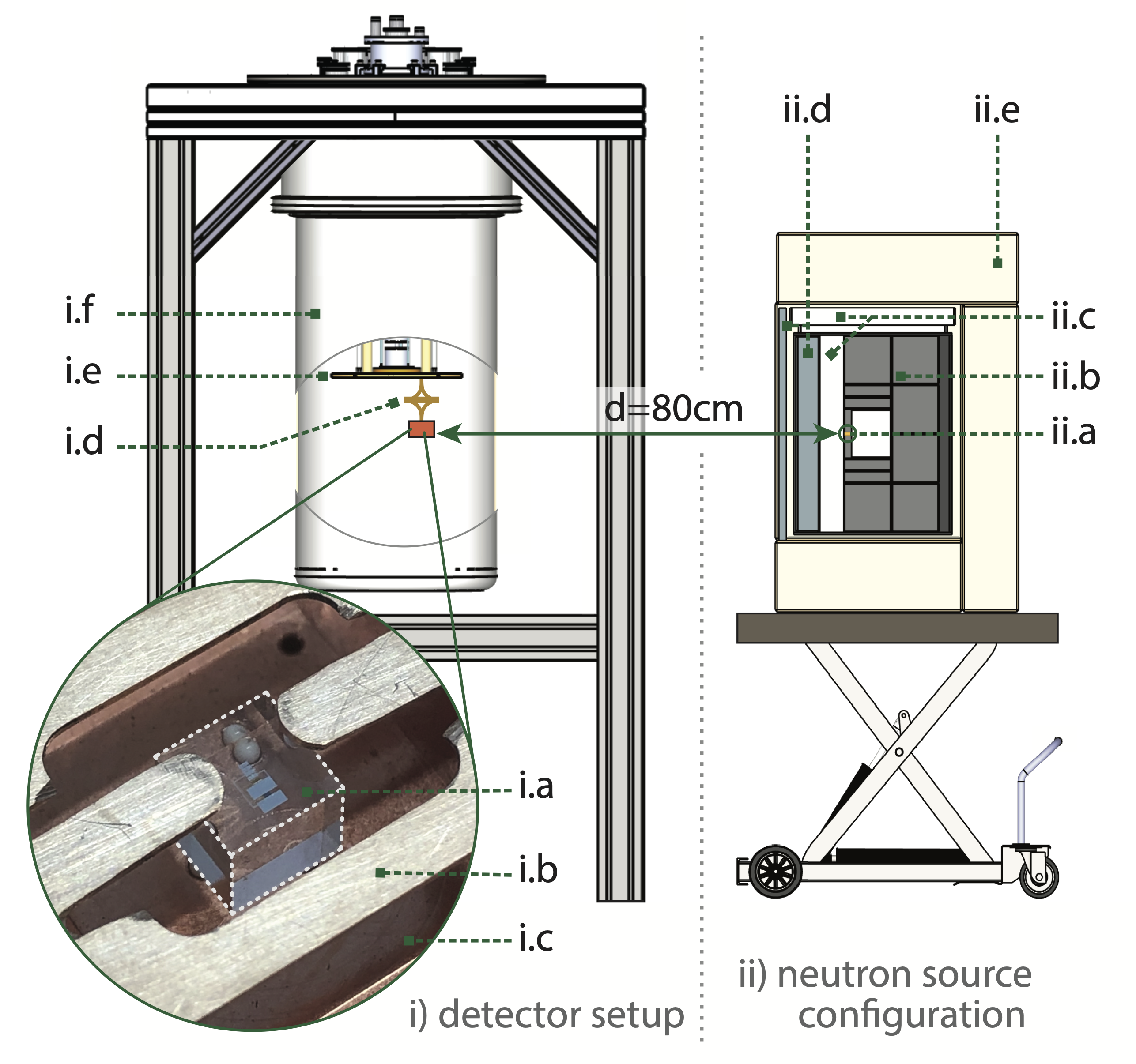}
    \caption{Overview of the cryostat with the detector setup $\textit{(i)}$ and the neutron source configuration $\textit{(ii)}$. The detector setup consists of a CaWO$_4$ cube equipped with a W thin-film TES $\textit{(i.a)}$ held by bronze clamps $\textit{(i.b)}$ inside a copper housing $\textit{(i.c)}$. The copper support box is mechanically decoupled via a two-stage spiral spring $\textit{(i.d)}$ from the mixing chamber $\textit{(i.e)}$ of the dry dilution refrigerator $\textit{(i.f)}$. The \nuc{252}{Cf} neutron source is inserted in a central PE cube $\textit{(ii.a)}$ surrounded by several layers of graphite $\textit{(ii.b)}$, PE $\textit{(ii.c)}$, lead $\textit{(ii.d)}$ and borated PE $\textit{(ii.e)}$.}
    \label{fig:Device}
\end{figure}

The experimental setup consists of a 0.75\,g cubic CaWO$_4$ crystal equipped with a tungsten thin-film Transition Edge Sensor (TES) \cite{Strauss:2017cuu}. The detector was built for the first phase of \NUCLEUS{} and is based on TES technology developed for the \CRESST{} experiment \cite{Angloher:2019flc}. 
It is operated in a dry dilution refrigerator (Bluefors LD400) with a base temperature of $<$\,7\,mK, located at the Technical University of Munich. The TES has a superconducting transition temperature of 20\,mK. 
It is biased with a constant current and stabilized in the operating point by a heater deposited close to the TES structure. 
The pulses are amplified and read out by a SQUID system and continuously streamed with a sampling rate of 400\,kHz and a precision of 20\,bits. The crystal is held between small sapphire spheres supported by flexible bronze clamps in a dedicated copper support box.
A thin gold wire provides the thermal connection between the crystal and the copper holder, which acts as a thermal bath. The detector assembly is decoupled mechanically from the mixing chamber by a two-stage spiral spring made from bronze. This efficiently attenuates vibrations induced by the operation of the pulse tube cooler, which was constantly in operation throughout the measurements.  An illustration of the experimental setup is shown in Fig.~\ref{fig:Device}.  
The detector is continuously exposed to an \nuc{55}{Fe} source providing the K$_{\alpha}$-lines at 5.985\,keV (weighted average) and the K$_\beta$ lines at 6.490\,keV of \nuc{}{Mn} for electronic recoil calibration.

The cryogenic detector was exposed to a flux of thermal neutrons originating from a commercial \nuc{252}{Cf} source with an activity of 3.54\,MBq positioned outside of the cryostat. \nuc{252}{Cf} decays by spontaneous fission with a branching ratio of 3\,\%, emitting on average 3.77~neutrons per fission with an average energy of 2.12\,MeV~\cite{Martin:2014you,Litaize:2015rco}. The other decay channel, $\alpha$ emission, has no impact on this measurement. A compact moderator was designed to fit inside a plexiglass box of dimensions (32$\times$34$\times$42)\,cm$^3$. At its center, the source capsule is placed inside a polyethylene (PE) cube of 10\,cm side length. In the direction of the cryostat, a 5\,cm thick layer of PE for neutron thermalization and a 7\,cm thick layer of lead to reduce the gamma flux 
are placed. Graphite blocks slow down and reflect neutrons and a layer of borated PE reduces the radiation dose in the surroundings (see Fig.~\ref{fig:Device} for details). 
This configuration --- verified by simulations (see later) ---  optimizes the flux of thermal neutrons emitted in the direction of the cryostat, while reducing the fast neutron and source-induced $\gamma$-ray background. 
The source setup is installed on a mobile lifting table. 
The neutron source was placed at a distance of $(80\,\pm\,1)$\,cm from the cryogenic detector, which results in a particle rate of 0.52\,cps. 
Two data sets were acquired: \textit{background data} to characterize the ambient background (lifetime 18.9\,h) and \textit{source data} with the \nuc{252}{Cf} source in place (lifetime 40.2\,h). 

A complete description of the detector, its vicinity including the full cryostat, and the neutron source have been implemented in a \GEANT \cite{GEANT4} Monte Carlo (MC) simulation. The \FIFRELIN{} code~\cite{Litaize:2015rco} was used to obtain an accurate prediction of the particles emitted by the fission of \nuc{252}{Cf} and by the de-excitation of tungsten nuclei following a neutron capture. From this simulation a thermal neutron rate of $\approx$0.25\,n$_{\text{th}}$/s is expected at the surface of the cryogenic detector leading to 114\,events in the 112.5\,eV peak for the \textit{source} run. 
To estimate the systematic uncertainty on the expected number of events, we take into account the uncertainty on the source activity (15\%)~\cite{ezag} and on the description of the geometry and materials in the MC simulations (20\%), leading to a total systematic uncertainty of 25\%  on the predicted number of counts in the peak. 

Two independent data analyses have been performed using two different frameworks:   DIANA~(analysis~1) developed for the CUORE experiments~\cite{CUORE:2016acf, Azzolini:2018yye} and adopted to the \NUCLEUS{} analysis~\cite{DelCastello::2021}, and CAT~(analysis~2) used e.g. by the CRESST experiment~\cite{Ferreiro:2019,Stahlberg:2020}.
In the following we present the results of analysis~1, being fully compatible with analysis~2 within uncertainties. To avoid any bias in the analysis event selection cuts were defined on the \textit{background data}.
In parallel, independent statistical tests were developed on simulated data.
 
\begin{figure}[t]
    \centering
    \includegraphics[width=1.0\linewidth]{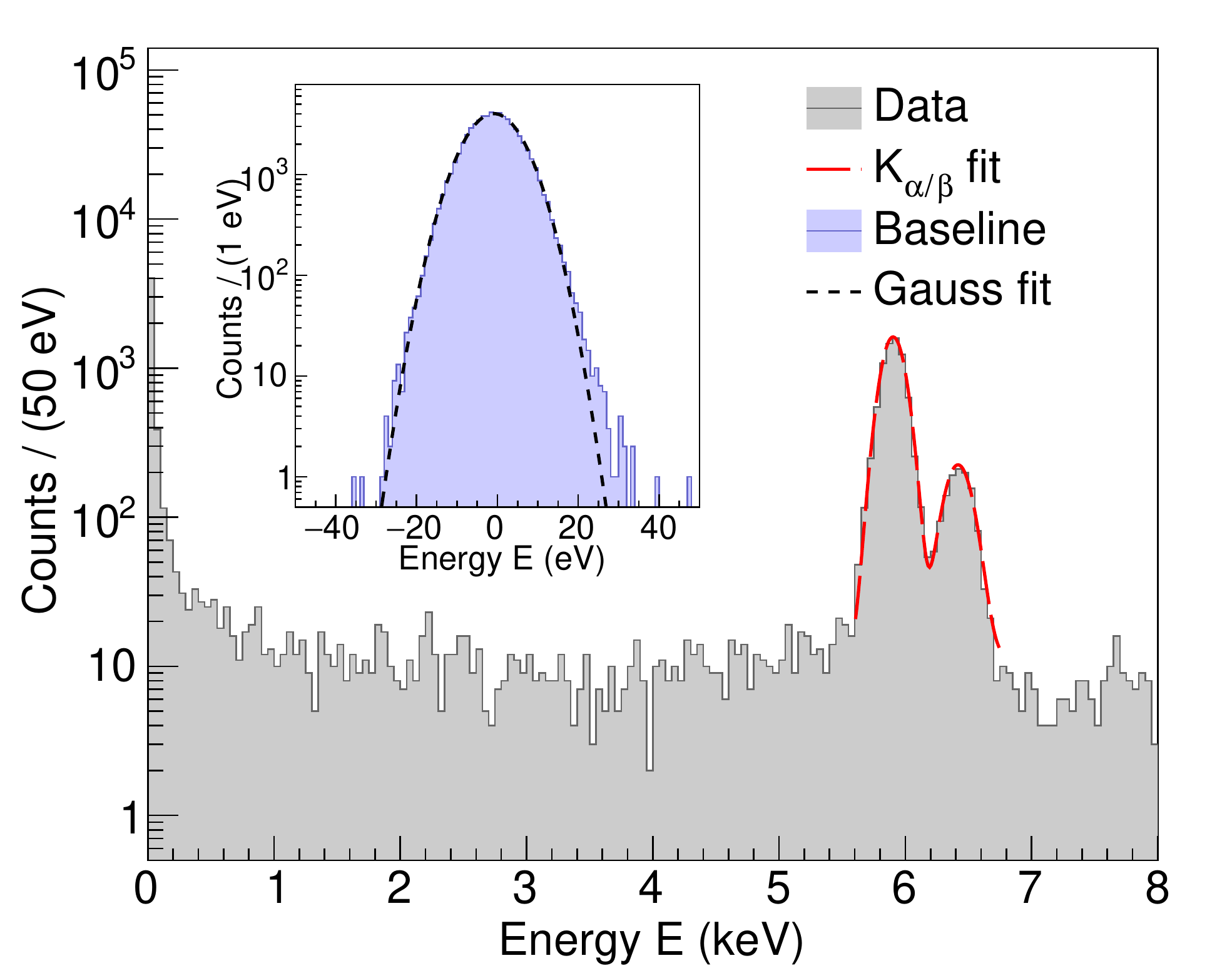}
    \caption{Energy spectrum recorded during the \textit{source} measurement. The two peaks from the \nuc{55}{Fe} source are used to set the energy scale of the detector (see text). The inset shows the distribution of filtered baselines after quality cuts. } 
    \label{fig:Exp_Fe}
\end{figure}

A software trigger is applied to the continuously acquired data stream using an optimum filter calculated from randomly chosen noise samples and a template built from particle pulses~\cite{DiDomizio:2010ph}.  
With a resolution of (94.2\,$\pm$\,1.4)\,eV in the 6\,keV energy range, the \nuc{}{Mn} K$_\alpha$ and K$_\beta$ lines can be clearly separated. Figure~\ref{fig:Exp_Fe} shows the reconstructed energy spectrum of the \textit{source data}, with a net \nuc{55}{Fe} rate of 0.13\,cps. 
The Mn K$_{\alpha/\beta}$ events lie well below the TES saturation level. Therefore, a simple linear extrapolation of the energy calibration towards lower pulse heights is assumed.  
Possible non-linearities in the detector response, originating, e.g. from the shape of the transition curve or the electrical readout circuit~\cite{CRESST:2020tlq}, are not compensated for in this analysis. These effects imply a rather large combined systematic uncertainty of [$-$18,+25]\,\% on the reconstructed energy at 112.5\,eV.

\begin{figure}[t]
    \centering
    \includegraphics[width=1.0\linewidth]{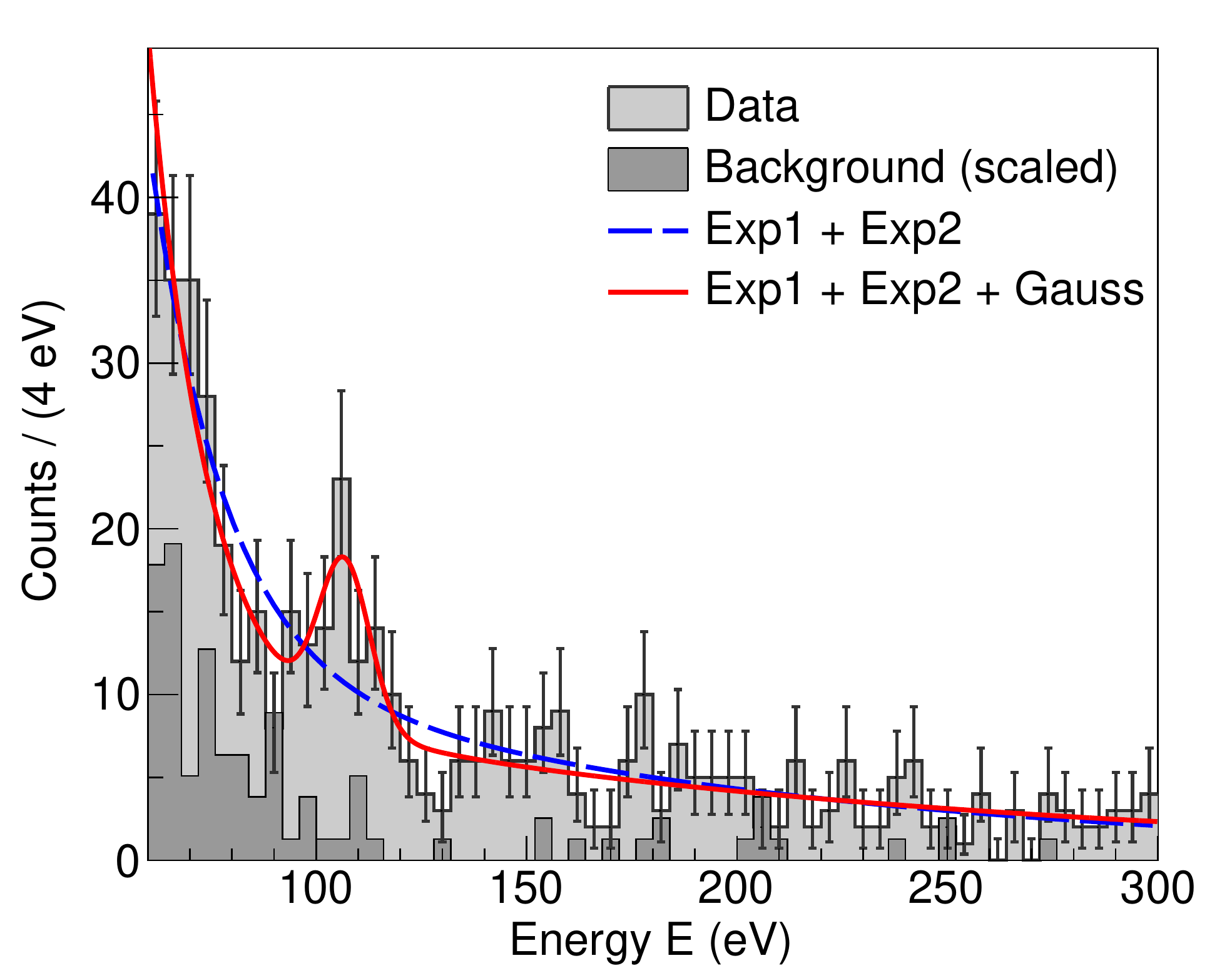}
    \caption{Energy spectra (from 60 to 300\,eV) measured by the \NUCLEUS{} detector for the \textit{source} and \textit{background} (scaled to source exposure) measurements. The error bars represent the Poissonian uncertainties. The red solid and blue dashed lines illustrate the best fit with and without the Gaussian contribution, respectively.} 
    \label{fig:Exp_Spectra}
\end{figure}

Quality cuts are used to clean the data from pulse artifacts and pile-up events.
The efficiency of the quality cuts is evaluated at zero energy with randomly chosen baseline samples and at 6\,keV using \nuc{55}{Fe} events. 
Reconstruction and trigger efficiencies are included in the final efficiency. We estimate a detection efficiency of ($46.3\pm8.3$)\,\% in the region-of-interest (60 -- 300 eV), given by the average of the efficiency evaluated at zero energy (40.4\,\%) and 6\,\,keV (52.2\,\%). 
The efficiency-corrected \nuc{55}{Fe} count rates match within 3\,\%  for \textit{source} and \textit{background data}, as well as within 7\,\% in between the two analysis frameworks, demonstrating the robustness of this approach. 
A baseline resolution of $\sigma_{\text{BL}}$\,=\,(6.37\,$\pm$\,0.02$_\text{(stat)}$)\,eV  [(6.54\,$\pm$\,0.02)\,eV] is observed for  \textit{background} [\textit{source}] measurement, see inset in Fig.~\ref{fig:Exp_Fe}. 
A conservative analysis threshold of 50\,eV is chosen corresponding to 8\,$\sigma_{\text{BL}}$. 
A total rate of reconstructed events of 0.5\,cps in the \textit{background data} and an increase by 0.2\,cps with the neutron source in place are measured. No significant drifts over time are observed neither on the position of the two \nuc{55}{Fe} peaks nor in the baseline resolutions. 
Figure \ref{fig:Exp_Spectra} shows the measured \textit{source} spectrum (light gray) compared to the \textit{background} spectrum (dark grey). 
The \textit{source} spectrum features a peak-like structure, centered at an energy compatible with the nominal calibration peak at 112.5\,eV above background.

In order to quantify the significance of this local structure, a simple statistical test has been developed. 
An effective model with two exponential functions describes the steep rise at low energy and the expected contribution of the scattering of fast neutrons from the source at higher energies.
A Gaussian describes the sought-for calibration peak. Two consecutive binned likelihood fits are then performed in the range $[60-300]$\,eV, with ($\mathcal{L}_{\text{Bck+Sig}}$) and without ($\mathcal{L}_{\text{Bck}}$) the contribution of the Gaussian (see solid red and dashed blue lines in Fig.~\ref{fig:Exp_Spectra})~\cite{Workman:2022ynf}.   

All parameters in the fits are left free to absorb possible experimental effects.  The $\chi^2$ distribution of the constructed statistical test t\,=\,$-2\ln{(\frac{\mathcal{L}_{\text{Bck}}}{\mathcal{L}_{\text{Bck+Sig}}})}$ has been simulated by generating numerous statistical realizations of our background model and was found to slightly deviate from the expected $\chi^2$ law for 3 degrees of freedom (d.o.f.). With t\,=\,14.86, the hypothesis of the non-existence of a Gaussian peak is rejected with 3.1\,$\sigma$ (2-sided). 
The best-fit yields a Gaussian peak with a mean value of $\mu_{\text{peak}}$\,=\,106.7$^{+1.9}_{-2.0}$\,eV, a standard deviation of $\sigma_{\text{peak}}$\,=\,6.0$^{+1.7}_{-1.4}$\,eV and an integral of 36.8$^{+9.7}_{-8.9}$\,counts. 
Correcting for the detector efficiency, the ratio of measured to predicted number of events in the peak is 0.70$\pm$0.29. The uncertainty on the ratio takes into account the systematic and statistical uncertainty on the predicted number of events (in total 27\%),  the detection efficiency (18\%), and the statistical uncertainty on the number of observed events.
The choice of the background model is validated a posteriori by the high p-value of the fit (0.82) when excluding the region $\mu_{\text{peak}}\,\pm\,3\sigma_{\text{peak}}$.
Analysis 2 finds compatible peak parameters and a statistical significance of 2.9\,$\sigma$.
Considering the simplified calibration based on a linear extrapolation of the detector response from 6\,keV down to threshold, and possible un-accounted non-linearities, the reconstructed peak is compatible with the nominal nuclear recoil peak at 112.5\,eV. 
The peak resolution agrees with the measured baseline detector resolution within uncertainties. 
Despite the low neutron flux and the large background from fast neutrons, the fit uncertainty on the mean position of the peak is already as low as 2\%(stat) in this short measurement, clearly opening the perspective for future precision studies. 

\begin{figure}[t]
    \centering
    \includegraphics[width=1.0\linewidth]{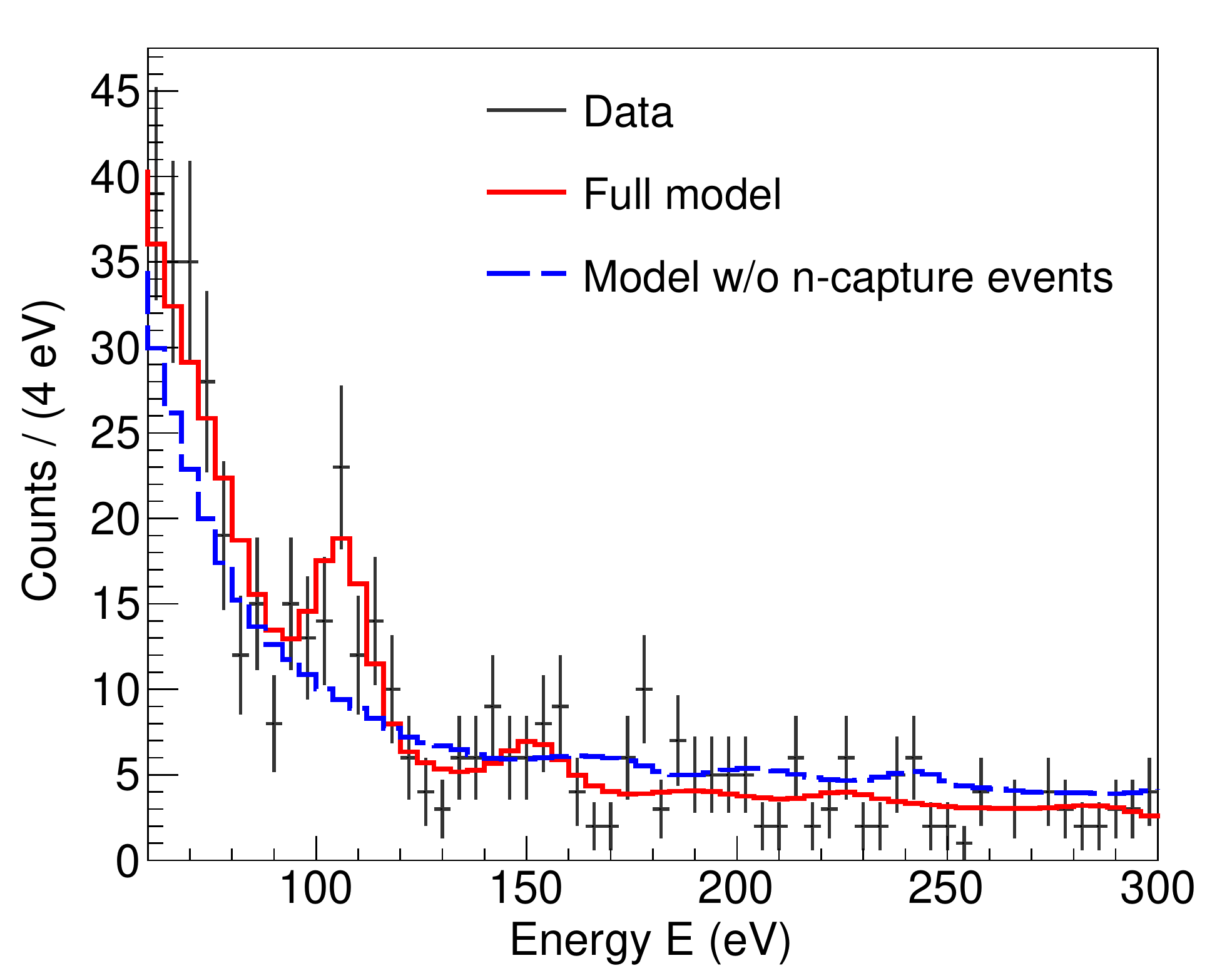}
    \caption{Comparison between the recoil spectrum measured with the neutron source in place (black points) and the expected spectrum built from the measured ambient background and the simulation with (red line curve) and without (blue curve) the predicted recoils induced by neutron capture. Error bars are statistical only. 
    }
    \label{fig:Data_Simu}
\end{figure}

As a complementary approach, we test the contribution of neutron captures in the measured spectrum. 
The expected recoil spectrum is predicted by \GEANT MC simulation. The latter can be used to separate the contributions of n-capture events (\textit{signal}) from other events (\textit{source background}) which in the 60 -- 300~eV range originate mainly from fast neutron scattering. 
The ambient background (which is not included in the simulation) is modeled with an exponential plus constant fit of the measured background spectrum, re-scaled to the lifetime of the source run. This contribution is kept fixed and added to the MC spectra.  
A statistical test similar to the previous peak search is thus defined based on two fits to the data, with and without the signal component. A linear correction of the energy scale, $\alpha$, and an energy-independent resolution, $\sigma$, are applied to all simulated events, and two independent normalization factors, $K_{bck}$ and $K_{sig}$, are applied to the \textit{source background} and the \textit{signal} predicted spectra, respectively. Thus, the statistical test is expected to follow a $\chi^2$ law for 1 d.o.f. from the $K_{sig}$ parameter. A MC study shows a slight deviation from this law, but again with little impact on the result. With a t-value of 42.3, the hypothesis of the non-existence of the n-captured-induced recoil spectrum is rejected with more than 6 $\sigma$ (2-sided). Figure~\ref{fig:Data_Simu} shows the comparison between the two best-fit models and the data. When including the signal recoil spectrum a very good agreement is observed, quantified by a $\chi^2/$ndf of 58.09/59. The parameters of the simulated spectrum converge to values fully compatible with those obtained in the peak search (see above): $\alpha=0.946\,\pm\,0.014$  is equivalent to locating the peak at (106.4\,$\pm$\,1.6)\,eV instead of the nominal 112.5 eV, and the fitted resolution is $\sigma=(6.00\,\pm\,0.47)$\,eV. Taking into account the systematic uncertainty on the predicted rate and the detection efficiency, the normalization $K_{sig}=0.74\,\pm\,0.28$ agrees with the ratio of measured to predicted number of events in the peak.

In the current data, the less prominent capture peak on \nuc{183}{W} at 160.3\,eV~\cite{Thulliez:2020esw} has low statistical significance, however, a future high-statistics measurement under the same experimental conditions  may be sensitive to this feature. 
Above $\sim120$\,eV, the background is dominated by fast neutrons from the source, a contribution that will be significantly suppressed when using a thermal neutron beam in the future~\cite{Thulliez:2020esw}.

The combined high significance of these statistical tests leads us to conclude that we have observed the first neutron-capture-induced peak at the 100\,eV scale with a CaWO$_4$ cryogenic detector of the \NUCLEUS{} experiment.
The position and resolution  of this peak are found to be compatible with the expectations for neutron capture on \nuc{182}{W} given the experimental uncertainties. 
The result represents a milestone for the calibration and understanding of low-threshold detectors, although the current measurement is limited by statistics and the lack of electron recoil calibration sources. Both will be improved in the near future with extensive measurements using low-energy X-rays and LED calibration sources~\cite{Cardani:2021iff}. Such measurements will enable the study of the phonon physics at lowest energies, e.g. the impact of crystal defect creation, as well as detector-related effects, and quenching between the electron and nuclear recoil energy scales.
Source-related background contributions can be significantly reduced by using a thermal neutron beam from a nuclear research reactor as proposed in~\cite{Thulliez:2020esw}, which also allows accessing a wider energy range.  
The calibration of most cryogenic detectors, e.g. those using Ge, Si, PbWO$_4$, and Al$_2$O$_3$, is in principle accessible by this new calibration method.
The demonstration of this method with a simple, portable neutron source enables an in-situ calibration of CE$\nu$NS and Dark Matter experiments in a non-intrusive manner --- a potential breakthrough for future low-threshold detectors being operated at the precision frontier.

\begin{acknowledgments}  
We thank M. Stahlberg from MPI for his support with the CAT analysis tool, which has been developed for the CRESST, COSINUS, and NUCLEUS experiments. This work made use of the DIANA software, originally developed for the CUORE and CUPID experiments and ported to the NUCLEUS experiment. 
This work has been funded by the DFG through the SFB1258. 
NUCLEUS is supported by the CEA, the INFN, the ÖAW, TU Wien, TU M\"unchen, and the MPI für Physik.  
We acknowledge additional funding by the DFG through the Excellence Cluster ORIGINS, 
by the European Commission through the ERC-StG2018-804228 ``NU-CLEUS”, by the P2IO LabEx (ANR-10-LABX-0038) in the 
framework ``Investissements d’Avenir" (ANR-11-IDEX-0003-01) managed by the Agence Nationale de la Recherche (ANR), France,
and by the Austrian Science Fund (FWF) through the projects ``I 5427-N CRAB" and ``P 34778-N ELOISE". 

\end{acknowledgments}

\bibliography{crab}

\providecommand{\noopsort}[1]{}\providecommand{\singleletter}[1]{#1}%
\begin{thebibliography}{42}%
\makeatletter
\providecommand \@ifxundefined [1]{%
 \@ifx{#1\undefined}
}%
\providecommand \@ifnum [1]{%
 \ifnum #1\expandafter \@firstoftwo
 \else \expandafter \@secondoftwo
 \fi
}%
\providecommand \@ifx [1]{%
 \ifx #1\expandafter \@firstoftwo
 \else \expandafter \@secondoftwo
 \fi
}%
\providecommand \natexlab [1]{#1}%
\providecommand \enquote  [1]{``#1''}%
\providecommand \bibnamefont  [1]{#1}%
\providecommand \bibfnamefont [1]{#1}%
\providecommand \citenamefont [1]{#1}%
\providecommand \href@noop [0]{\@secondoftwo}%
\providecommand \href [0]{\begingroup \@sanitize@url \@href}%
\providecommand \@href[1]{\@@startlink{#1}\@@href}%
\providecommand \@@href[1]{\endgroup#1\@@endlink}%
\providecommand \@sanitize@url [0]{\catcode `\\12\catcode `\$12\catcode
  `\&12\catcode `\#12\catcode `\^12\catcode `\_12\catcode `\%12\relax}%
\providecommand \@@startlink[1]{}%
\providecommand \@@endlink[0]{}%
\providecommand \url  [0]{\begingroup\@sanitize@url \@url }%
\providecommand \@url [1]{\endgroup\@href {#1}{\urlprefix }}%
\providecommand \urlprefix  [0]{URL }%
\providecommand \Eprint [0]{\href }%
\providecommand \doibase [0]{https://doi.org/}%
\providecommand \selectlanguage [0]{\@gobble}%
\providecommand \bibinfo  [0]{\@secondoftwo}%
\providecommand \bibfield  [0]{\@secondoftwo}%
\providecommand \translation [1]{[#1]}%
\providecommand \BibitemOpen [0]{}%
\providecommand \bibitemStop [0]{}%
\providecommand \bibitemNoStop [0]{.\EOS\space}%
\providecommand \EOS [0]{\spacefactor3000\relax}%
\providecommand \BibitemShut  [1]{\csname bibitem#1\endcsname}%
\let\auto@bib@innerbib\@empty
\bibitem [{\citenamefont {Strauss}\ \emph
  {et~al.}(2017{\natexlab{a}})\citenamefont {Strauss} \emph
  {et~al.}}]{Strauss:2017cam}%
  \BibitemOpen
  \bibfield  {author} {\bibinfo {author} {\bibfnamefont {R.}~\bibnamefont
  {Strauss}} \emph {et~al.},\ }\bibfield  {title} {\bibinfo {title}
  {{Gram-scale cryogenic calorimeters for rare-event searches}},\ }\href
  {https://doi.org/10.1103/PhysRevD.96.022009} {\bibfield  {journal} {\bibinfo
  {journal} {Phys. Rev. D}\ }\textbf {\bibinfo {volume} {96}},\ \bibinfo
  {pages} {022009} (\bibinfo {year} {2017}{\natexlab{a}})},\ \Eprint
  {https://arxiv.org/abs/1704.04317} {arXiv:1704.04317 [physics.ins-det]}
  \BibitemShut {NoStop}%
\bibitem [{\citenamefont {Abdelhameed}\ \emph {et~al.}(2019)\citenamefont
  {Abdelhameed} \emph {et~al.}}]{CRESST:2019jnq}%
  \BibitemOpen
  \bibfield  {author} {\bibinfo {author} {\bibfnamefont {A.~H.}\ \bibnamefont
  {Abdelhameed}} \emph {et~al.} (\bibinfo {collaboration} {CRESST}),\
  }\bibfield  {title} {\bibinfo {title} {{First results from the CRESST-III
  low-mass dark matter program}},\ }\href
  {https://doi.org/10.1103/PhysRevD.100.102002} {\bibfield  {journal} {\bibinfo
   {journal} {Phys. Rev. D}\ }\textbf {\bibinfo {volume} {100}},\ \bibinfo
  {pages} {102002} (\bibinfo {year} {2019})},\ \Eprint
  {https://arxiv.org/abs/1904.00498} {arXiv:1904.00498 [astro-ph.CO]}
  \BibitemShut {NoStop}%
\bibitem [{\citenamefont {Armengaud}\ \emph {et~al.}(2019)\citenamefont
  {Armengaud} \emph {et~al.}}]{PhysRevD.99.082003}%
  \BibitemOpen
  \bibfield  {author} {\bibinfo {author} {\bibfnamefont {E.}~\bibnamefont
  {Armengaud}} \emph {et~al.} (\bibinfo {collaboration} {EDELWEISS
  Collaboration}),\ }\bibfield  {title} {\bibinfo {title} {{Searching for
  low-mass dark matter particles with a massive Ge bolometer operated above
  ground}},\ }\href {https://doi.org/10.1103/PhysRevD.99.082003} {\bibfield
  {journal} {\bibinfo  {journal} {Phys. Rev. D}\ }\textbf {\bibinfo {volume}
  {99}},\ \bibinfo {pages} {082003} (\bibinfo {year} {2019})},\ \Eprint
  {https://arxiv.org/abs/1901.03588} {arXiv:1901.03588 [astro-ph.GA]}
  \BibitemShut {NoStop}%
\bibitem [{\citenamefont {Armengaud}\ \emph {et~al.}(2022)\citenamefont
  {Armengaud} \emph {et~al.}}]{EDELWEISS:2022ktt}%
  \BibitemOpen
  \bibfield  {author} {\bibinfo {author} {\bibfnamefont {E.}~\bibnamefont
  {Armengaud}} \emph {et~al.} (\bibinfo {collaboration} {EDELWEISS}),\
  }\bibfield  {title} {\bibinfo {title} {{Search for sub-GeV dark matter via
  the Migdal effect with an EDELWEISS germanium detector with NbSi
  transition-edge sensors}},\ }\href
  {https://doi.org/10.1103/PhysRevD.106.062004} {\bibfield  {journal} {\bibinfo
   {journal} {Phys. Rev. D}\ }\textbf {\bibinfo {volume} {106}},\ \bibinfo
  {pages} {062004} (\bibinfo {year} {2022})},\ \Eprint
  {https://arxiv.org/abs/2203.03993} {arXiv:2203.03993 [astro-ph.GA]}
  \BibitemShut {NoStop}%
\bibitem [{\citenamefont {Alkhatib}\ \emph {et~al.}(2021)\citenamefont
  {Alkhatib} \emph {et~al.}}]{SuperCDMS:2020aus}%
  \BibitemOpen
  \bibfield  {author} {\bibinfo {author} {\bibfnamefont {I.}~\bibnamefont
  {Alkhatib}} \emph {et~al.} (\bibinfo {collaboration} {SuperCDMS}),\
  }\bibfield  {title} {\bibinfo {title} {{Light Dark Matter Search with a
  High-Resolution Athermal Phonon Detector Operated Above Ground}},\ }\href
  {https://doi.org/10.1103/PhysRevLett.127.061801} {\bibfield  {journal}
  {\bibinfo  {journal} {Phys. Rev. Lett.}\ }\textbf {\bibinfo {volume} {127}},\
  \bibinfo {pages} {061801} (\bibinfo {year} {2021})},\ \Eprint
  {https://arxiv.org/abs/2007.14289} {arXiv:2007.14289 [hep-ex]} \BibitemShut
  {NoStop}%
\bibitem [{\citenamefont {Agnese}\ \emph {et~al.}(2018)\citenamefont {Agnese}
  \emph {et~al.}}]{SuperCDMS:2018mne}%
  \BibitemOpen
  \bibfield  {author} {\bibinfo {author} {\bibfnamefont {R.}~\bibnamefont
  {Agnese}} \emph {et~al.} (\bibinfo {collaboration} {SuperCDMS}),\ }\bibfield
  {title} {\bibinfo {title} {{First Dark Matter Constraints from a SuperCDMS
  Single-Charge Sensitive Detector}},\ }\href
  {https://doi.org/10.1103/PhysRevLett.121.051301} {\bibfield  {journal}
  {\bibinfo  {journal} {{Phys. Rev. Lett.}}\ }\textbf {\bibinfo {volume}
  {121}},\ \bibinfo {pages} {051301} (\bibinfo {year} {2018})},\ \bibinfo
  {note} {[Erratum: Phys. Rev. Lett. 122, 069901 (2019)]},\ \Eprint
  {https://arxiv.org/abs/1804.10697} {arXiv:1804.10697 [hep-ex]} \BibitemShut
  {NoStop}%
\bibitem [{\citenamefont {Angloher}\ \emph {et~al.}(2017)\citenamefont
  {Angloher} \emph {et~al.}}]{Angloher:2017sxg}%
  \BibitemOpen
  \bibfield  {author} {\bibinfo {author} {\bibfnamefont {G.}~\bibnamefont
  {Angloher}} \emph {et~al.} (\bibinfo {collaboration} {CRESST}),\ }\bibfield
  {title} {\bibinfo {title} {{Results on MeV-scale dark matter from a
  gram-scale cryogenic calorimeter operated above ground}},\ }\href
  {https://doi.org/10.1140/epjc/s10052-017-5223-9} {\bibfield  {journal}
  {\bibinfo  {journal} {Eur. Phys. J. C}\ }\textbf {\bibinfo {volume} {77}},\
  \bibinfo {pages} {637} (\bibinfo {year} {2017})},\ \Eprint
  {https://arxiv.org/abs/1707.06749} {arXiv:1707.06749 [astro-ph.CO]}
  \BibitemShut {NoStop}%
\bibitem [{\citenamefont {Freedman}(1974)}]{Freedman:1973yd}%
  \BibitemOpen
  \bibfield  {author} {\bibinfo {author} {\bibfnamefont {D.~Z.}\ \bibnamefont
  {Freedman}},\ }\bibfield  {title} {\bibinfo {title} {{Coherent Neutrino
  Nucleus Scattering as a Probe of the Weak Neutral Current}},\ }\href
  {https://doi.org/10.1103/PhysRevD.9.1389} {\bibfield  {journal} {\bibinfo
  {journal} {Phys.\ Rev.\ D}\ }\textbf {\bibinfo {volume} {9}},\ \bibinfo
  {pages} {1389} (\bibinfo {year} {1974})}\BibitemShut {NoStop}%
\bibitem [{\citenamefont {Drukier}\ and\ \citenamefont
  {Stodolsky}(1984)}]{Drukier:1983gj}%
  \BibitemOpen
  \bibfield  {author} {\bibinfo {author} {\bibfnamefont {A.}~\bibnamefont
  {Drukier}}\ and\ \bibinfo {author} {\bibfnamefont {L.}~\bibnamefont
  {Stodolsky}},\ }\bibfield  {title} {\bibinfo {title} {{Principles and
  Applications of a Neutral Current Detector for Neutrino Physics and
  Astronomy}},\ }\href {https://doi.org/10.1103/PhysRevD.30.2295} {\bibfield
  {journal} {\bibinfo  {journal} {Phys. Rev. D}\ }\textbf {\bibinfo {volume}
  {30}},\ \bibinfo {pages} {2295} (\bibinfo {year} {1984})}\BibitemShut
  {NoStop}%
\bibitem [{\citenamefont {Abdullah}\ \emph {et~al.}(2022)\citenamefont
  {Abdullah} \emph {et~al.}}]{Abdullah:2022zue}%
  \BibitemOpen
  \bibfield  {author} {\bibinfo {author} {\bibfnamefont {M.}~\bibnamefont
  {Abdullah}} \emph {et~al.},\ }\bibfield  {title} {\bibinfo {title} {{Coherent
  elastic neutrino-nucleus scattering: Terrestrial and astrophysical
  applications}},\ }\href@noop {} {\  (\bibinfo {year} {2022})},\ \Eprint
  {https://arxiv.org/abs/2203.07361} {arXiv:2203.07361 [hep-ph]} \BibitemShut
  {NoStop}%
\bibitem [{\citenamefont {Aristizabal~Sierra}\ \emph
  {et~al.}(2021)\citenamefont {Aristizabal~Sierra}, \citenamefont {Dutta},
  \citenamefont {Kim}, \citenamefont {Snowden-Ifft},\ and\ \citenamefont
  {Strigari}}]{AristizabalSierra:2021uob}%
  \BibitemOpen
  \bibfield  {author} {\bibinfo {author} {\bibfnamefont {D.}~\bibnamefont
  {Aristizabal~Sierra}}, \bibinfo {author} {\bibfnamefont {B.}~\bibnamefont
  {Dutta}}, \bibinfo {author} {\bibfnamefont {D.}~\bibnamefont {Kim}}, \bibinfo
  {author} {\bibfnamefont {D.}~\bibnamefont {Snowden-Ifft}},\ and\ \bibinfo
  {author} {\bibfnamefont {L.~E.}\ \bibnamefont {Strigari}},\ }\bibfield
  {title} {\bibinfo {title} {{Coherent elastic neutrino-nucleus scattering with
  the \ensuremath{\nu}BDX-DRIFT directional detector at next generation
  neutrino facilities}},\ }\href {https://doi.org/10.1103/PhysRevD.104.033004}
  {\bibfield  {journal} {\bibinfo  {journal} {Phys. Rev. D}\ }\textbf {\bibinfo
  {volume} {104}},\ \bibinfo {pages} {033004} (\bibinfo {year} {2021})},\
  \Eprint {https://arxiv.org/abs/2103.10857} {arXiv:2103.10857 [hep-ph]}
  \BibitemShut {NoStop}%
\bibitem [{\citenamefont {Lindner}\ \emph {et~al.}(2017)\citenamefont
  {Lindner}, \citenamefont {Rodejohann},\ and\ \citenamefont
  {Xu}}]{Lindner:2016wff}%
  \BibitemOpen
  \bibfield  {author} {\bibinfo {author} {\bibfnamefont {M.}~\bibnamefont
  {Lindner}}, \bibinfo {author} {\bibfnamefont {W.}~\bibnamefont
  {Rodejohann}},\ and\ \bibinfo {author} {\bibfnamefont {X.-J.}\ \bibnamefont
  {Xu}},\ }\bibfield  {title} {\bibinfo {title} {{Coherent Neutrino-Nucleus
  Scattering and new Neutrino Interactions}},\ }\href
  {https://doi.org/10.1007/JHEP03(2017)097} {\bibfield  {journal} {\bibinfo
  {journal} {{J. High Energ. Phys. \bf{2017}}}\ ,\ \bibinfo {pages} {097}}
  (\bibinfo {year} {2017})},\ \Eprint {https://arxiv.org/abs/1612.04150}
  {arXiv:1612.04150 [hep-ph]} \BibitemShut {NoStop}%
\bibitem [{\citenamefont {Dent}\ \emph {et~al.}(2017)\citenamefont {Dent},
  \citenamefont {Dutta}, \citenamefont {Liao}, \citenamefont {Newstead},
  \citenamefont {Strigari},\ and\ \citenamefont {Walker}}]{Dent:2016wcr}%
  \BibitemOpen
  \bibfield  {author} {\bibinfo {author} {\bibfnamefont {J.~B.}\ \bibnamefont
  {Dent}}, \bibinfo {author} {\bibfnamefont {B.}~\bibnamefont {Dutta}},
  \bibinfo {author} {\bibfnamefont {S.}~\bibnamefont {Liao}}, \bibinfo {author}
  {\bibfnamefont {J.~L.}\ \bibnamefont {Newstead}}, \bibinfo {author}
  {\bibfnamefont {L.~E.}\ \bibnamefont {Strigari}},\ and\ \bibinfo {author}
  {\bibfnamefont {J.~W.}\ \bibnamefont {Walker}},\ }\bibfield  {title}
  {\bibinfo {title} {{Probing light mediators at ultralow threshold energies
  with coherent elastic neutrino-nucleus scattering}},\ }\href
  {https://doi.org/10.1103/PhysRevD.96.095007} {\bibfield  {journal} {\bibinfo
  {journal} {Phys. Rev. D}\ }\textbf {\bibinfo {volume} {96}},\ \bibinfo
  {pages} {095007} (\bibinfo {year} {2017})},\ \Eprint
  {https://arxiv.org/abs/1612.06350} {arXiv:1612.06350 [hep-ph]} \BibitemShut
  {NoStop}%
\bibitem [{\citenamefont {Strauss}\ \emph
  {et~al.}(2017{\natexlab{b}})\citenamefont {Strauss} \emph
  {et~al.}}]{Strauss:2017cuu}%
  \BibitemOpen
  \bibfield  {author} {\bibinfo {author} {\bibfnamefont {R.}~\bibnamefont
  {Strauss}} \emph {et~al.},\ }\bibfield  {title} {\bibinfo {title} {{The
  $\nu$-cleus experiment: A gram-scale fiducial-volume cryogenic detector for
  the first detection of coherent neutrino-nucleus scattering}},\ }\href
  {https://doi.org/10.1140/epjc/s10052-017-5068-2} {\bibfield  {journal}
  {\bibinfo  {journal} {Eur. Phys. J. C}\ }\textbf {\bibinfo {volume} {77}},\
  \bibinfo {pages} {506} (\bibinfo {year} {2017}{\natexlab{b}})},\ \Eprint
  {https://arxiv.org/abs/1704.04320} {arXiv:1704.04320 [physics.ins-det]}
  \BibitemShut {NoStop}%
\bibitem [{\citenamefont {Rothe}\ \emph {et~al.}(2019)\citenamefont {Rothe}
  \emph {et~al.}}]{Rothe:2019aii}%
  \BibitemOpen
  \bibfield  {author} {\bibinfo {author} {\bibfnamefont {J.}~\bibnamefont
  {Rothe}} \emph {et~al.} (\bibinfo {collaboration} {NUCLEUS}),\ }\bibfield
  {title} {\bibinfo {title} {{NUCLEUS: Exploring Coherent Neutrino-Nucleus
  Scattering with Cryogenic Detectors}},\ }\href
  {https://doi.org/10.1007/s10909-019-02283-7} {\bibfield  {journal} {\bibinfo
  {journal} {J. Low Temp. Phys.}\ }\textbf {\bibinfo {volume} {199}},\ \bibinfo
  {pages} {433} (\bibinfo {year} {2019})}\BibitemShut {NoStop}%
\bibitem [{\citenamefont {Angloher}\ \emph {et~al.}(2019)\citenamefont
  {Angloher} \emph {et~al.}}]{Angloher:2019flc}%
  \BibitemOpen
  \bibfield  {author} {\bibinfo {author} {\bibfnamefont {G.}~\bibnamefont
  {Angloher}} \emph {et~al.} (\bibinfo {collaboration} {NUCLEUS}),\ }\bibfield
  {title} {\bibinfo {title} {{Exploring $\hbox {CE}\nu \hbox {NS}$ with NUCLEUS
  at the Chooz nuclear power plant}},\ }\href
  {https://doi.org/10.1140/epjc/s10052-019-7454-4} {\bibfield  {journal}
  {\bibinfo  {journal} {Eur. Phys. J. C}\ }\textbf {\bibinfo {volume} {79}},\
  \bibinfo {pages} {1018} (\bibinfo {year} {2019})},\ \Eprint
  {https://arxiv.org/abs/1905.10258} {arXiv:1905.10258 [physics.ins-det]}
  \BibitemShut {NoStop}%
\bibitem [{\citenamefont {Augier}\ \emph {et~al.}(2021)\citenamefont {Augier}
  \emph {et~al.}}]{Ricochet:2021rjo}%
  \BibitemOpen
  \bibfield  {author} {\bibinfo {author} {\bibfnamefont {C.}~\bibnamefont
  {Augier}} \emph {et~al.} (\bibinfo {collaboration} {Ricochet}),\ }\bibfield
  {title} {\bibinfo {title} {{Ricochet Progress and Status}},\ }in\ \href@noop
  {} {\emph {\bibinfo {booktitle} {{19th International Workshop on Low
  Temperature Detectors}}}}\ (\bibinfo {year} {2021})\ \Eprint
  {https://arxiv.org/abs/2111.06745} {arXiv:2111.06745 [physics.ins-det]}
  \BibitemShut {NoStop}%
\bibitem [{\citenamefont {Agnolet}\ \emph {et~al.}(2017)\citenamefont {Agnolet}
  \emph {et~al.}}]{Agnolet:2016zir}%
  \BibitemOpen
  \bibfield  {author} {\bibinfo {author} {\bibfnamefont {G.}~\bibnamefont
  {Agnolet}} \emph {et~al.} (\bibinfo {collaboration} {MINER}),\ }\bibfield
  {title} {\bibinfo {title} {{Background Studies for the MINER Coherent
  Neutrino Scattering Reactor Experiment}},\ }\href
  {https://doi.org/10.1016/j.nima.2017.02.024} {\bibfield  {journal} {\bibinfo
  {journal} {Nucl. Instrum. Meth. A}\ }\textbf {\bibinfo {volume} {853}},\
  \bibinfo {pages} {53} (\bibinfo {year} {2017})},\ \Eprint
  {https://arxiv.org/abs/1609.02066} {arXiv:1609.02066 [physics.ins-det]}
  \BibitemShut {NoStop}%
\bibitem [{\citenamefont {Bonet}\ \emph {et~al.}(2021)\citenamefont {Bonet}
  \emph {et~al.}}]{bonet2021constraints}%
  \BibitemOpen
  \bibfield  {author} {\bibinfo {author} {\bibfnamefont {H.}~\bibnamefont
  {Bonet}} \emph {et~al.} (\bibinfo {collaboration} {CONUS}),\ }\bibfield
  {title} {\bibinfo {title} {{Constraints on Elastic Neutrino Nucleus
  Scattering in the Fully Coherent Regime from the CONUS Experiment}},\ }\href
  {https://doi.org/10.1103/PhysRevLett.126.041804} {\bibfield  {journal}
  {\bibinfo  {journal} {Phys. Rev. Lett.}\ }\textbf {\bibinfo {volume} {126}},\
  \bibinfo {pages} {041804} (\bibinfo {year} {2021})},\ \Eprint
  {https://arxiv.org/abs/2011.00210} {arXiv:2011.00210 [hep-ex]} \BibitemShut
  {NoStop}%
\bibitem [{\citenamefont {Aguilar-Arevalo}\ \emph {et~al.}(2020)\citenamefont
  {Aguilar-Arevalo} \emph {et~al.}}]{CONNIE:2019xid}%
  \BibitemOpen
  \bibfield  {author} {\bibinfo {author} {\bibfnamefont {A.}~\bibnamefont
  {Aguilar-Arevalo}} \emph {et~al.} (\bibinfo {collaboration} {CONNIE}),\
  }\bibfield  {title} {\bibinfo {title} {{Search for light mediators in the
  low-energy data of the CONNIE reactor neutrino experiment}},\ }\href
  {https://doi.org/10.1007/JHEP04(2020)054} {\bibfield  {journal} {\bibinfo
  {journal} {{J. High Energ. Phys. \bf{2020}}}\ ,\ \bibinfo {pages} {054}}
  (\bibinfo {year} {2020})},\ \Eprint {https://arxiv.org/abs/1910.04951}
  {arXiv:1910.04951 [hep-ex]} \BibitemShut {NoStop}%
\bibitem [{\citenamefont {Akimov}\ \emph {et~al.}(2017)\citenamefont {Akimov}
  \emph {et~al.}}]{Akimov:2017hee}%
  \BibitemOpen
  \bibfield  {author} {\bibinfo {author} {\bibfnamefont {D.~Y.}\ \bibnamefont
  {Akimov}} \emph {et~al.},\ }\bibfield  {title} {\bibinfo {title} {{Status of
  the RED-100 experiment}},\ }\href
  {https://doi.org/10.1088/1748-0221/12/06/C06018} {\bibfield  {journal}
  {\bibinfo  {journal} {JINST}\ }\textbf {\bibinfo {volume} {12}},\ \bibinfo
  {pages} {C06018 (2017)}}\BibitemShut {NoStop}%
\bibitem [{\citenamefont {Ramanathan}\ and\ \citenamefont
  {Kurinsky}(2020)}]{PhysRevD.102.063026}%
  \BibitemOpen
  \bibfield  {author} {\bibinfo {author} {\bibfnamefont {K.}~\bibnamefont
  {Ramanathan}}\ and\ \bibinfo {author} {\bibfnamefont {N.}~\bibnamefont
  {Kurinsky}},\ }\bibfield  {title} {\bibinfo {title} {Ionization yield in
  silicon for ev-scale electron-recoil processes},\ }\href
  {https://doi.org/10.1103/PhysRevD.102.063026} {\bibfield  {journal} {\bibinfo
   {journal} {Phys. Rev. D}\ }\textbf {\bibinfo {volume} {102}},\ \bibinfo
  {pages} {063026} (\bibinfo {year} {2020})}\BibitemShut {NoStop}%
\bibitem [{\citenamefont {Baxter}\ \emph {et~al.}(2022)\citenamefont {Baxter}
  \emph {et~al.}}]{Baxter:2022dkm}%
  \BibitemOpen
  \bibfield  {author} {\bibinfo {author} {\bibfnamefont {D.}~\bibnamefont
  {Baxter}} \emph {et~al.},\ }\bibfield  {title} {\bibinfo {title}
  {{Snowmass2021 Cosmic Frontier White Paper: Calibrations and backgrounds for
  dark matter direct detection}},\ }\href@noop {} {\  (\bibinfo {year}
  {2022})},\ \Eprint {https://arxiv.org/abs/2203.07623} {arXiv:2203.07623
  [hep-ex]} \BibitemShut {NoStop}%
\bibitem [{\citenamefont {Sassi}\ \emph {et~al.}(2022)\citenamefont {Sassi},
  \citenamefont {Heikinheimo}, \citenamefont {Tuominen}, \citenamefont
  {Kuronen}, \citenamefont {Byggm\"astar}, \citenamefont {Nordlund},\ and\
  \citenamefont {Mirabolfathi}}]{PhysRevD.106.063012}%
  \BibitemOpen
  \bibfield  {author} {\bibinfo {author} {\bibfnamefont {S.}~\bibnamefont
  {Sassi}}, \bibinfo {author} {\bibfnamefont {M.}~\bibnamefont {Heikinheimo}},
  \bibinfo {author} {\bibfnamefont {K.}~\bibnamefont {Tuominen}}, \bibinfo
  {author} {\bibfnamefont {A.}~\bibnamefont {Kuronen}}, \bibinfo {author}
  {\bibfnamefont {J.}~\bibnamefont {Byggm\"astar}}, \bibinfo {author}
  {\bibfnamefont {K.}~\bibnamefont {Nordlund}},\ and\ \bibinfo {author}
  {\bibfnamefont {N.}~\bibnamefont {Mirabolfathi}},\ }\bibfield  {title}
  {\bibinfo {title} {Energy loss in low energy nuclear recoils in dark matter
  detector materials},\ }\href {https://doi.org/10.1103/PhysRevD.106.063012}
  {\bibfield  {journal} {\bibinfo  {journal} {Phys. Rev. D}\ }\textbf {\bibinfo
  {volume} {106}},\ \bibinfo {pages} {063012} (\bibinfo {year}
  {2022})}\BibitemShut {NoStop}%
\bibitem [{\citenamefont {Thulliez}\ \emph {et~al.}(2021)\citenamefont
  {Thulliez} \emph {et~al.}}]{Thulliez:2020esw}%
  \BibitemOpen
  \bibfield  {author} {\bibinfo {author} {\bibfnamefont {L.}~\bibnamefont
  {Thulliez}} \emph {et~al.},\ }\bibfield  {title} {\bibinfo {title}
  {{Calibration of nuclear recoils at the 100 eV scale using neutron
  capture}},\ }\href {https://doi.org/10.1088/1748-0221/16/07/P07032}
  {\bibfield  {journal} {\bibinfo  {journal} {JINST}\ }\textbf {\bibinfo
  {volume} {16}},\ \bibinfo {pages} {P07032 (2021)}},\ \Eprint
  {https://arxiv.org/abs/2011.13803} {arXiv:2011.13803 [physics.ins-det]}
  \BibitemShut {NoStop}%
\bibitem [{\citenamefont {Villano}\ \emph {et~al.}(2022)\citenamefont
  {Villano}, \citenamefont {Fritts}, \citenamefont {Mast}, \citenamefont
  {Brown}, \citenamefont {Cushman}, \citenamefont {Harris},\ and\ \citenamefont
  {Mandic}}]{Villano:2021eof}%
  \BibitemOpen
  \bibfield  {author} {\bibinfo {author} {\bibfnamefont {A.~N.}\ \bibnamefont
  {Villano}}, \bibinfo {author} {\bibfnamefont {M.}~\bibnamefont {Fritts}},
  \bibinfo {author} {\bibfnamefont {N.}~\bibnamefont {Mast}}, \bibinfo {author}
  {\bibfnamefont {S.}~\bibnamefont {Brown}}, \bibinfo {author} {\bibfnamefont
  {P.}~\bibnamefont {Cushman}}, \bibinfo {author} {\bibfnamefont
  {K.}~\bibnamefont {Harris}},\ and\ \bibinfo {author} {\bibfnamefont
  {V.}~\bibnamefont {Mandic}},\ }\bibfield  {title} {\bibinfo {title} {{First
  observation of isolated nuclear recoils following neutron capture for dark
  matter calibration}},\ }\href {https://doi.org/10.1103/PhysRevD.105.083014}
  {\bibfield  {journal} {\bibinfo  {journal} {Phys. Rev. D}\ }\textbf {\bibinfo
  {volume} {105}},\ \bibinfo {pages} {083014} (\bibinfo {year} {2022})},\
  \Eprint {https://arxiv.org/abs/2110.02751} {arXiv:2110.02751 [nucl-ex]}
  \BibitemShut {NoStop}%
\bibitem [{iso()}]{isoAbundance}%
  \BibitemOpen
  \href@noop {} {\bibinfo {title} {{Commission on isotopic abundances and
  atomic weights}}},\ \bibinfo {howpublished}
  {{\url{https://www.ciaaw.org/isotopic-abundances.htm}}},\ \bibinfo {note}
  {accessed: 2022-11-03}\BibitemShut {NoStop}%
\bibitem [{NIS(2022)}]{NISTxcom}%
  \BibitemOpen
  \href@noop {} {\bibinfo {title} {{NIST} xcom, data base for photon
  cross-section}},\ \bibinfo {howpublished}
  {\url{https://physics.nist.gov/PhysRefData/Xcom/html/xcom1.html}} (\bibinfo
  {year} {2022}),\ \bibinfo {note} {accessed: 2022-11-03}\BibitemShut {NoStop}%
\bibitem [{\citenamefont {Brown}\ \emph {et~al.}(2018)\citenamefont {Brown}
  \emph {et~al.}}]{Brown:2018jhj}%
  \BibitemOpen
  \bibfield  {author} {\bibinfo {author} {\bibfnamefont {D.~A.}\ \bibnamefont
  {Brown}} \emph {et~al.},\ }\bibfield  {title} {\bibinfo {title}
  {{ENDF/B-VIII.0: The 8th Major Release of the Nuclear Reaction Data Library
  with CIELO-project Cross Sections, New Standards and Thermal Scattering
  Data}},\ }\href {https://doi.org/10.1016/j.nds.2018.02.001} {\bibfield
  {journal} {\bibinfo  {journal} {Nucl. Data Sheets}\ }\textbf {\bibinfo
  {volume} {148}},\ \bibinfo {pages} {1} (\bibinfo {year} {2018})}\BibitemShut
  {NoStop}%
\bibitem [{\citenamefont {Martin}(2014)}]{Martin:2014you}%
  \BibitemOpen
  \bibfield  {author} {\bibinfo {author} {\bibfnamefont {M.~J.}\ \bibnamefont
  {Martin}},\ }\bibfield  {title} {\bibinfo {title} {{Nuclear Data Sheets for A
  = 248}},\ }\href {https://doi.org/10.1016/j.nds.2014.11.004} {\bibfield
  {journal} {\bibinfo  {journal} {Nucl. Data Sheets}\ }\textbf {\bibinfo
  {volume} {122}},\ \bibinfo {pages} {377} (\bibinfo {year}
  {2014})}\BibitemShut {NoStop}%
\bibitem [{\citenamefont {Litaize}\ \emph {et~al.}(2015)\citenamefont
  {Litaize}, \citenamefont {Serot},\ and\ \citenamefont
  {Berge}}]{Litaize:2015rco}%
  \BibitemOpen
  \bibfield  {author} {\bibinfo {author} {\bibfnamefont {O.}~\bibnamefont
  {Litaize}}, \bibinfo {author} {\bibfnamefont {O.}~\bibnamefont {Serot}},\
  and\ \bibinfo {author} {\bibfnamefont {L.}~\bibnamefont {Berge}},\ }\bibfield
   {title} {\bibinfo {title} {{Fission modelling with FIFRELIN}},\ }\href
  {https://doi.org/10.1140/epja/i2015-15177-9} {\bibfield  {journal} {\bibinfo
  {journal} {Eur. Phys. J. A}\ }\textbf {\bibinfo {volume} {51}},\ \bibinfo
  {pages} {177} (\bibinfo {year} {2015})}\BibitemShut {NoStop}%
\bibitem [{\citenamefont {Allison}\ \emph {et~al.}(2006)\citenamefont {Allison}
  \emph {et~al.}}]{GEANT4}%
  \BibitemOpen
  \bibfield  {author} {\bibinfo {author} {\bibnamefont {Allison}} \emph
  {et~al.},\ }\bibfield  {title} {\bibinfo {title} {Geant4 developments and
  applications},\ }\href {https://doi.org/10.1109/TNS.2006.869826} {\bibfield
  {journal} {\bibinfo  {journal} {IEEE Trans. Nucl. Sci.}\ }\textbf {\bibinfo
  {volume} {53}},\ \bibinfo {pages} {270} (\bibinfo {year} {2006})}\BibitemShut
  {NoStop}%
\bibitem [{eza()}]{ezag}%
  \BibitemOpen
  \href@noop {} {\bibinfo {title} {{Eckert~\&~Ziegler} reference and
  calibration sources, product information}},\ \bibinfo {howpublished}
  {\url{https://www.ezag.com/fileadmin/user_upload/isotopes/isotopes/Isotrak/isotrak-pdf/Product_literature/EZIPL/EZIP_catalogue_reference_and_calibration_sources.pdf}},\
  \bibinfo {note} {accessed: 2022-11-03}\BibitemShut {NoStop}%
\bibitem [{\citenamefont {Alduino}\ \emph {et~al.}(2016)\citenamefont {Alduino}
  \emph {et~al.}}]{CUORE:2016acf}%
  \BibitemOpen
  \bibfield  {author} {\bibinfo {author} {\bibfnamefont {C.}~\bibnamefont
  {Alduino}} \emph {et~al.} (\bibinfo {collaboration} {CUORE}),\ }\bibfield
  {title} {\bibinfo {title} {{Analysis techniques for the evaluation of the
  neutrinoless double-$\beta$ decay lifetime in $^{130}$Te with the CUORE-0
  detector}},\ }\href {https://doi.org/10.1103/PhysRevC.93.045503} {\bibfield
  {journal} {\bibinfo  {journal} {Phys. Rev. C}\ }\textbf {\bibinfo {volume}
  {93}},\ \bibinfo {pages} {045503} (\bibinfo {year} {2016})},\ \Eprint
  {https://arxiv.org/abs/1601.01334} {arXiv:1601.01334 [nucl-ex]} \BibitemShut
  {NoStop}%
\bibitem [{\citenamefont {Azzolini}\ \emph {et~al.}(2018)\citenamefont
  {Azzolini} \emph {et~al.}}]{Azzolini:2018yye}%
  \BibitemOpen
  \bibfield  {author} {\bibinfo {author} {\bibfnamefont {O.}~\bibnamefont
  {Azzolini}} \emph {et~al.},\ }\bibfield  {title} {\bibinfo {title} {{Analysis
  of cryogenic calorimeters with light and heat read-out for double beta decay
  searches}},\ }\href {https://doi.org/10.1140/epjc/s10052-018-6202-5}
  {\bibfield  {journal} {\bibinfo  {journal} {Eur. Phys. J. C}\ }\textbf
  {\bibinfo {volume} {78}},\ \bibinfo {pages} {734} (\bibinfo {year} {2018})},\
  \Eprint {https://arxiv.org/abs/1806.02826} {arXiv:1806.02826
  [physics.ins-det]} \BibitemShut {NoStop}%
\bibitem [{\citenamefont {Del~Castello}(2021)}]{DelCastello::2021}%
  \BibitemOpen
  \bibfield  {author} {\bibinfo {author} {\bibfnamefont {G.}~\bibnamefont
  {Del~Castello}},\ }\emph {\bibinfo {title} {Development of energy calibration
  and data analysis system for the NUCLEUS experiment}},\ \href
  {https://doi.org/10.48550/arXiv.2302.02843} {Master's thesis},\ \bibinfo
  {school} {Sapienza -- University of Rome} (\bibinfo {year}
  {2021})\BibitemShut {NoStop}%
\bibitem [{\citenamefont {Ferreiro~Iachellini}(2019)}]{Ferreiro:2019}%
  \BibitemOpen
  \bibfield  {author} {\bibinfo {author} {\bibfnamefont {N.}~\bibnamefont
  {Ferreiro~Iachellini}},\ }\emph {\bibinfo {title} {{Increasing the
  sensitivity to low mass dark matter in CRESST-III with a new DAQ and signal
  processing}}},\ \href {10.5282/edoc.23762} {Ph.D. thesis},\ \bibinfo
  {school} {LMU München} (\bibinfo {year} {2019})\BibitemShut {NoStop}%
\bibitem [{\citenamefont {Stahlberg}(2020)}]{Stahlberg:2020}%
  \BibitemOpen
  \bibfield  {author} {\bibinfo {author} {\bibfnamefont {M.}~\bibnamefont
  {Stahlberg}},\ }\emph {\bibinfo {title} {{Probing low-mass dark matter with
  CRESST-III: data analysis and first results}}},\ \href
  {https://doi.org/10.34726/hss.2021.45935} {Ph.D. thesis},\ \bibinfo  {school}
  {TU Wien} (\bibinfo {year} {2020})\BibitemShut {NoStop}%
\bibitem [{\citenamefont {Di~Domizio}\ \emph {et~al.}(2011)\citenamefont
  {Di~Domizio}, \citenamefont {Orio},\ and\ \citenamefont
  {Vignati}}]{DiDomizio:2010ph}%
  \BibitemOpen
  \bibfield  {author} {\bibinfo {author} {\bibfnamefont {S.}~\bibnamefont
  {Di~Domizio}}, \bibinfo {author} {\bibfnamefont {F.}~\bibnamefont {Orio}},\
  and\ \bibinfo {author} {\bibfnamefont {M.}~\bibnamefont {Vignati}},\
  }\bibfield  {title} {\bibinfo {title} {{Lowering the energy threshold of
  large-mass bolometric detectors}},\ }\href
  {https://doi.org/10.1088/1748-0221/6/02/P02007} {\bibfield  {journal}
  {\bibinfo  {journal} {JINST}\ }\textbf {\bibinfo {volume} {6}},\ \bibinfo
  {pages} {P02007 (2011)}},\ \Eprint {https://arxiv.org/abs/1012.1263}
  {arXiv:1012.1263 [astro-ph.IM]} \BibitemShut {NoStop}%
\bibitem [{\citenamefont {Abdelhameed}\ \emph {et~al.}(2020)\citenamefont
  {Abdelhameed} \emph {et~al.}}]{CRESST:2020tlq}%
  \BibitemOpen
  \bibfield  {author} {\bibinfo {author} {\bibfnamefont {A.~H.}\ \bibnamefont
  {Abdelhameed}} \emph {et~al.} (\bibinfo {collaboration} {CRESST}),\
  }\bibfield  {title} {\bibinfo {title} {{Cryogenic characterization of a
  $\hbox {LiAlO}_{2}$ crystal and new results on spin-dependent dark matter
  interactions with ordinary matter}},\ }\href
  {https://doi.org/10.1140/epjc/s10052-020-8329-4} {\bibfield  {journal}
  {\bibinfo  {journal} {Eur. Phys. J. C}\ }\textbf {\bibinfo {volume} {80}},\
  \bibinfo {pages} {834} (\bibinfo {year} {2020})},\ \Eprint
  {https://arxiv.org/abs/2005.02692} {arXiv:2005.02692 [physics.ins-det]}
  \BibitemShut {NoStop}%
\bibitem [{\citenamefont {Workman}\ and\ \citenamefont
  {Others}(2022)}]{Workman:2022ynf}%
  \BibitemOpen
  \bibfield  {author} {\bibinfo {author} {\bibfnamefont {R.~L.}\ \bibnamefont
  {Workman}}\ and\ \bibinfo {author} {\bibnamefont {Others}} (\bibinfo
  {collaboration} {Particle Data Group}),\ }\bibfield  {title} {\bibinfo
  {title} {{Review of Particle Physics}},\ }\href
  {https://doi.org/10.1093/ptep/ptac097} {\bibfield  {journal} {\bibinfo
  {journal} {Prog. Theor. Exp. Phys.}\ }\textbf {\bibinfo {volume} {2022}},\
  \bibinfo {pages} {083C01} (\bibinfo {year} {2022})}\BibitemShut {NoStop}%
\bibitem [{\citenamefont {Cardani}\ \emph {et~al.}(2021)\citenamefont
  {Cardani}, \citenamefont {Casali}, \citenamefont {Colantoni}, \citenamefont
  {Cruciani}, \citenamefont {Di~Domizio}, \citenamefont {Martinez},
  \citenamefont {Pettinacci}, \citenamefont {Pettinari},\ and\ \citenamefont
  {Vignati}}]{Cardani:2021iff}%
  \BibitemOpen
  \bibfield  {author} {\bibinfo {author} {\bibfnamefont {L.}~\bibnamefont
  {Cardani}}, \bibinfo {author} {\bibfnamefont {N.}~\bibnamefont {Casali}},
  \bibinfo {author} {\bibfnamefont {I.}~\bibnamefont {Colantoni}}, \bibinfo
  {author} {\bibfnamefont {A.}~\bibnamefont {Cruciani}}, \bibinfo {author}
  {\bibfnamefont {S.}~\bibnamefont {Di~Domizio}}, \bibinfo {author}
  {\bibfnamefont {M.}~\bibnamefont {Martinez}}, \bibinfo {author}
  {\bibfnamefont {V.}~\bibnamefont {Pettinacci}}, \bibinfo {author}
  {\bibfnamefont {G.}~\bibnamefont {Pettinari}},\ and\ \bibinfo {author}
  {\bibfnamefont {M.}~\bibnamefont {Vignati}},\ }\bibfield  {title} {\bibinfo
  {title} {{Final results of CALDER: kinetic inductance light detectors to
  search for rare events}},\ }\href
  {https://doi.org/10.1140/epjc/s10052-021-09454-5} {\bibfield  {journal}
  {\bibinfo  {journal} {Eur. Phys. J. C}\ }\textbf {\bibinfo {volume} {81}},\
  \bibinfo {pages} {636} (\bibinfo {year} {2021})},\ \Eprint
  {https://arxiv.org/abs/2104.06850} {arXiv:2104.06850 [physics.ins-det]}
  \BibitemShut {NoStop}%
\end{thebibliography}%

\end{document}